\documentclass[aps,prb,amsmath,twocolumn,amssymb,amsbsy, titlepage]{revtex4-1}

\usepackage{graphicx}
\usepackage{nicefrac}
\usepackage{amsfonts}

\usepackage{amssymb}
\usepackage{amsmath} 

\usepackage{subfigure}
\usepackage{multirow} 
\usepackage{tabularx} 
\usepackage{array}

\usepackage{color}

\usepackage{units}

\usepackage{tensor} 
\usepackage{braket}

\usepackage{bm}
\usepackage{hyperref}

\begin{document}
\title{Collective enhancements in many-emitter phonon lasing}
\author{Leon Droenner}
\author{Nicolas L. Naumann}
\author{Julia Kabuss}

\author{Alexander Carmele} 
\affiliation{Institut f\"ur Theoretische Physik, Nichtlineare Optik und Quantenelektronik, Technische Universit\"at Berlin, 10623 Berlin, Germany}

% Repeat:
% \author{}
% \affiliation{}
% For each author

\begin{abstract}
We investigate theoretically the many-emitter phonon laser based on optically driven semiconductor quantum dots within an acoustic nanocavity. We map the phonon laser Hamiltonian to a Tavis-Cummings type interaction with an unexpected additional many-emitter energy shift. This many-emitter interaction with the cavity mode results in a variety of resonances dependent on the number of participating emitters. We show that the many-emitter phonon laser also includes the single emitter resonance besides these collective phenomena. However, we obtain a high quantum yield addressing these collective resonances. We clearly demonstrate the best setup for maximal enhancement and show that the output can be increased even more via lasing at the two phonon resonance.
\end{abstract}

\maketitle
%\section{Introduction}
The optical laser is indispensable for fundamental physics and many applications and is well understood by now \cite{Haken:Laser, weng:laser}. Adapting the concept of coherent amplification by stimulated emission to sound waves, the phonon laser is a promising candidate for a new type of non demolishing measurement \cite{viewpoint}. In the past few years there have been different theoretical and experimental proposals for phonon lasing such as trapped ions\cite{10.1038/nphys1367, 0295-5075-91-3-33001}, compound microcavities\cite{PhysRevLett.104.083901}, NV-centers\cite{PhysRevB.88.064105}, electromagnetic resonators \cite{PhysRevLett.110.127202} and semiconductor devices \cite{PhysRevB.64.125311, PhysRevLett.90.077402, PhysRevLett.104.085501, PhysRevLett.109.054301, kabuss:effectivehamiltonian, Naumann:16}. For most applications, especially for the latter, the embedding of the active medium within an acoustic cavity forms the basis for stimulated phonon emission. The design and technological control of such cavities have been progressing in the past years \cite{PhysRevLett.89.227402, PhysRevB.75.024301, LanzillottiKimura201580, PhysRevLett.102.015502, PhysRevLett.107.235502, PhysRevLett.110.037403}. In a semiconductor device, two superlattices confine one single phonon mode within a spacer in between\cite{PhysRevLett.89.227402, PhysRevB.75.024301, LanzillottiKimura201580}. A careful design allows long phonon lifetimes due to a high quality factor  up to $Q=10^5$ of the acoustic nanocavity\cite{PhysRevLett.102.015502, PhysRevLett.107.235502, PhysRevLett.110.037403}. Based on such a phonon cavity, we investigate theoretically phonon lasing in a semiconductor nanodevice with quantum dots \cite{PhysRevLett.105.157401} as active medium and external coherent optical excitation \cite{PhysRevB.73.035302} for coherent phonon generation via the induced raman process\cite{PhysRevLett.109.054301, kabuss:effectivehamiltonian}. \\ 
We want to extend the single emitter case \cite{PhysRevLett.109.054301} to many emitters  \cite{2017arXiv170104209W, PhysRevLett.118.133901} and focus on collective effects of the phonon lasing regime. In analogy to the Tavis-cummings model \cite{PhysRev.170.379, PhysRevB.91.035306} we focus on identical emitters coupled via the cavity phonon field. First discovered by Dicke \cite{PhysRev.93.99}, superradiance is one example for collective effects with applications to optical lasers \cite{PhysRevApplied.4.044018, PhysRevA.92.063832, PhysRevLett.110.113604} and has been investigated recently for phonons in general \cite{PhysRevLett.112.023603, PhysRevA.95.023806}. The coherent pump of identical emitters supports the buildup of collective quantum coherences \cite{PhysRevLett.112.023603, ANDP:ANDP201400144, 2017arXiv170502889G}. We clarify that collective phenomena also appear in the phonon lasing regime and show an enhancement of the coherent cavity phonon field addressing collective processes. Even if there are similarities to the optical laser, the phonon laser differs fundamentally in the interaction as the electron-phonon coupling is diagonal. This results in a Tavis-Cummings type interaction with an additional new type of interaction between the emitters via the cavity field.\\ This paper is structured as follows: In Sec. \ref{sec_model}, we explain our model system of optically driven semiconductor quantum dots (QDs) within an acoustic nanocavity and the scenario leading to coherent phonon generation. In Sec. \ref{sec:Collective two phonon process} we show that several resonances appear in the many-emitter phonon laser in addition to the single emitter resonance. This proves an unexpected robustness of the phonon laser against many-emitter effects. In order to explain these additional resonances, we derive an Tavis-Cummings like effective Hamiltonian which demonstrates a collective phonon emission process of different emitters at the specific resonances. Additionally, we demonstrate that a detuning at the two phonon resonance leads to a coherent two phonon generation of one single QD as well as a collective two phonon process including all QDs at lower frequencies. In Sec. \ref{nonidentical emitters} we investigate the case of non identical emitters. We give an estimation what difference in electronic transition frequencies or electron-phonon coupling can be tolerated in order to still observe collective phonon emission.\\
Finally, we investigate the quantum yield in Sec. \ref{sec:Collective quantum yield enhancement}.  Hereby, we show for a given scenario the best pumping strength and lasing frequency of the external laser, leading to maximal many-emitter enhancement.

\section{Model}\label{sec_model}
\begin{figure*}
\vspace{1cm}
\center
\flushleft{(a)\hspace{12cm}(b)}
\subfigure{\includegraphics[width=0.6\textwidth]{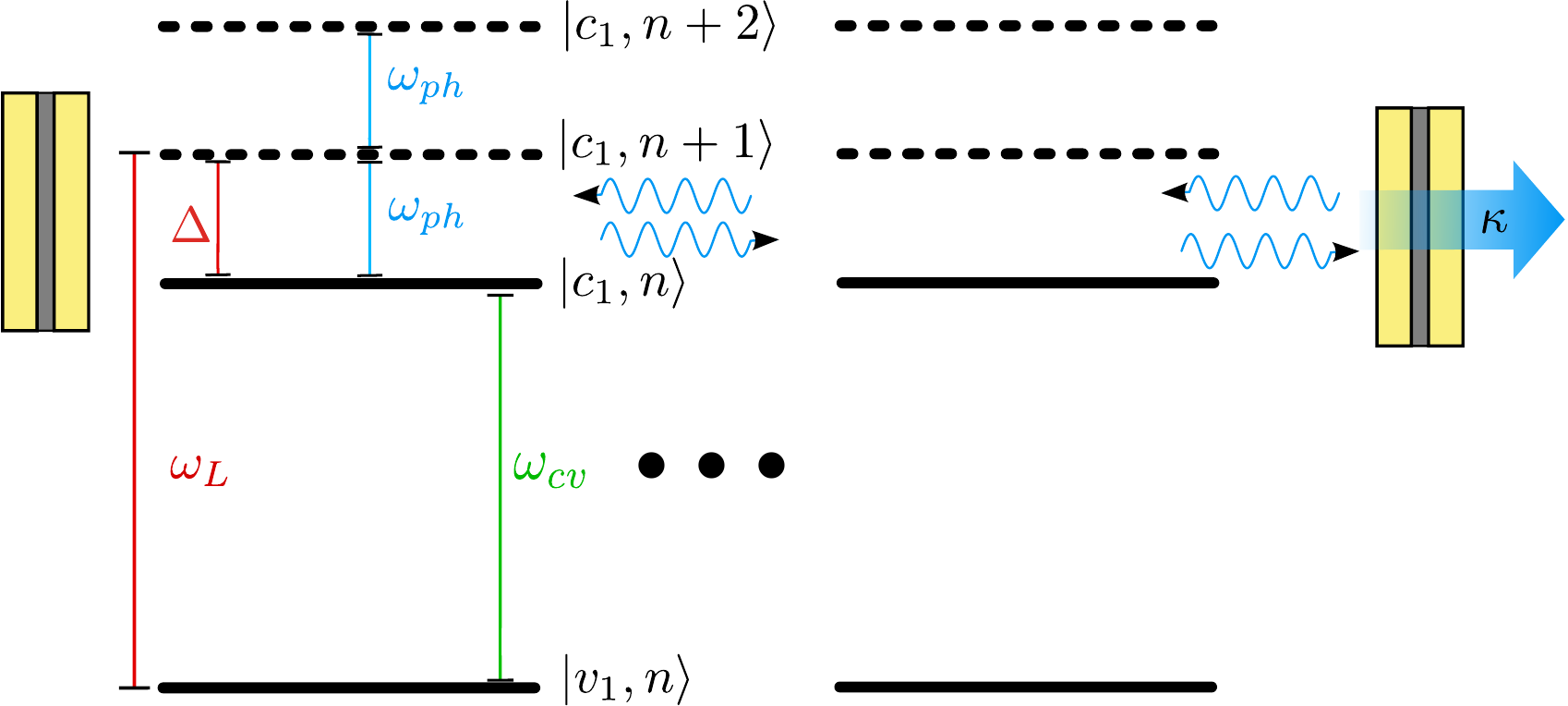}}
\hspace{1.5cm}
\subfigure{\includegraphics[width=0.25\textwidth]{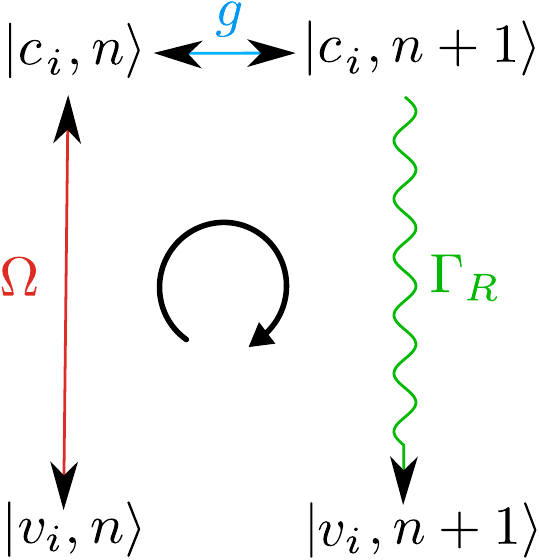}}
\caption{Illustration of the many-emitter phonon laser. (a) Energetic excitation scheme: Several QDs are placed inside an acoustic cavity with phonon frequency $\omega_{ph}$ and cavity loss $\kappa$. The external laser frequency $\omega_L$ is blue detuned from the two-level resonance $\omega_{cv}$ with respect to the cavity phonon frequency $\Delta\approx\omega_{ph}$ at the anti-Stokes resonance. (b) Phonon lasing cycle: First, the QDs are pumped coherently via Rabi frequency $\Omega$ from $|v_i,n\rangle$ to $|c_i,n\rangle$. Second, a phonon is emitted into the cavity via electron-phonon coupling $g$.  A radiative decay $\Gamma_R$ of the upper electronic level $|c_i,n+1\rangle$ closes the loop. Note that $\Gamma_R$ prevents the QD from reabsorbing phonons and the lasing cycle starts again.}
\label{fig:model-system}
\end{figure*} 
\begin{figure}[h!]
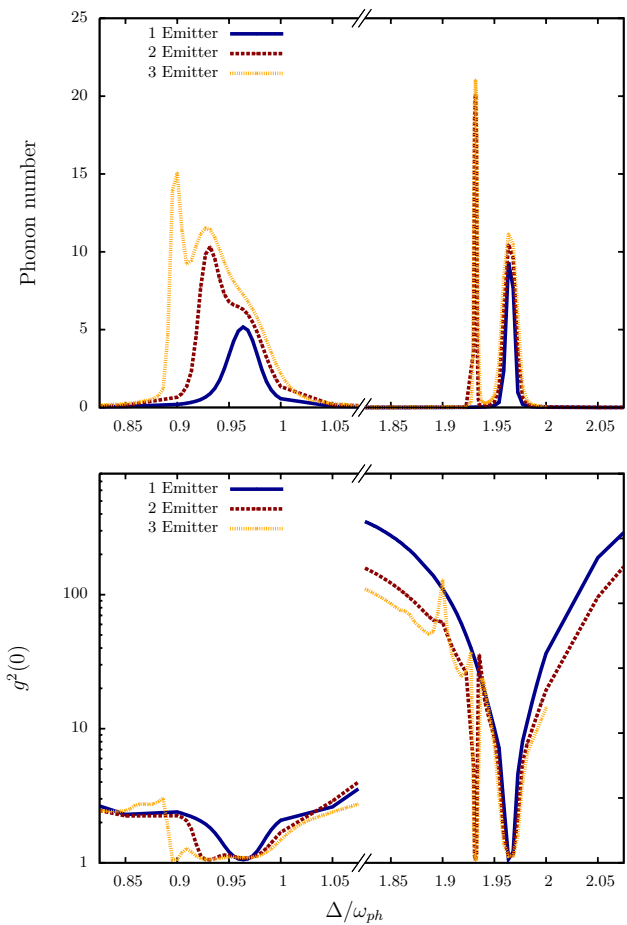

\centering
\resizebox{0.48\textwidth}{!}{\input{Phononnumber}}
\resizebox{0.48\textwidth}{!}{\input{g2}}
\caption{Optical detuning versus phonon number and $g^2(0)$-function For one emitter there is a single resonance close to the cavity frequency $\omega_{ph}$ and a second at the doubled frequency. Increasing the number of emitters we find as much resonances as there are emitters close to $\Delta\approx \omega_{ph}$. At each resonance the second-order correlation function shows $g^2(0)=1$ which signifies coherent phonon statistics. Parameters: $\omega_{cv}=2.28~ 1/fs$, $\omega_{ph}=0.011~ 1/fs$, $\Omega=4.56\cdot 10^{-4}~ 1/fs$, $g=2\cdot 10^{-3}~ 1/fs$, $\Gamma_R=1\cdot 10^{-5}~ 1/fs$, $\kappa=5\cdot 10^{-7}~ 1/fs$.} 
\label{figure:spectra}
\end{figure}
As active medium for coherent cavity phonon generation, we use semiconductor quantum dots, described as two-level systems with band-gap frequency $\omega_{cv}$, valence band state $|v\rangle_i$ and conduction band state $|c\rangle_i$. The quantum dots are embedded within a high-Q acoustic cavity with one single phonon mode $\omega_{ph}$. We drive the quantum dots coherently with an external optical laser with frequency $\omega_L$ and Rabi-frequency $\Omega(t)$. The optical pump, we describe semiclassically within rotating wave approximation due to quasi resonance with the two-level system and small pumping strengths compared to the driving frequency $\omega_L$. In order to generate phonon emission, the external optical laser is blue detuned at the anti-Stokes frequency $\Delta= \omega_L-\omega_{cv}\approx\omega_{ph}$. For simplicity, we transform the Hamiltonian into a rotating frame with respect to the excitation laser frequency $\omega_L$. The full Hamiltonian $\mathcal{H}=\mathcal{H}_0+\mathcal{H}_I$ reads
\begin{align}
\mathcal{H}_0&=\frac{\hbar \Delta}{2} \sum_{i=1}^{N_{QD}}\sigma^i_z+\hbar\omega_{ph}b^\dagger b~,\label{eq:H_0}\\
\mathcal{H}_I&=\sum_{i=1}^{N_{QD}}\left[ \hbar g \sigma^{i}_+\sigma^i_- (b^\dagger+b)+\hbar
\Omega (t) \sigma^i_x\right] ~,
\label{eq:H_I}
\end{align}
with $b^{(\dagger)}$ being the phonon annihilation (creation) operator and $\sigma^i_z=|v\rangle_{ii}\langle v|-|c\rangle_{ii}\langle c|$, $\sigma^i_x= |v\rangle_{ii}\langle c|+|c\rangle_{ii}\langle v|$ and $\sigma^i_-=|v\rangle_{ii}\langle c|$ the typical Pauli matrices.
Initially, we assume the system to be in a low temperature regime at $T=4K$. We neglect the back action from the phononic and the photonic reservoirs in order to assume the losses to be Markovian. Therefore, we describe the phonon decay of the cavity $\kappa$ and the radiative decay $\Gamma_R$ of the electronic excited states via the Lindblad master equation
\begin{align}
\label{eq:master}
\dot{\rho} = -\frac{i}{\hbar}[\mathcal{H}(t),\rho] + 2\kappa \mathcal{D}[b]\rho
+ 2\Gamma_R  \sum_{i=1}^{N_{QD}} \mathcal{D}[\sigma^i_-]\rho, 
\end{align} 
with super operator $ \mathcal{D}[x]\rho\equiv x\rho x^\dagger -\frac{1}{2}\{ x^\dagger x ,\rho
\}$.\\
In Fig. \ref{fig:model-system} we illustrate the many emitter phonon laser scheme.  The lasing cycle works as follows: The electron is pumped coherently via $\Omega$ from $|v_i,n\rangle$ to $|c_i,n\rangle$. Due to the excess in energy, a phonon is created via the electron-phonon interaction $g$, bringing the system to the state $|c_i,n+1\rangle$. After decaying radiatively to $|v_i,n+1\rangle$ with $\Gamma_R$, the electron may once again be excited. In Eq. \eqref{eq:H_0} and \eqref{eq:H_I} we have assumed identical emitters which are coupled only via the cavity mode $\omega_{ph}$.\\
In the Refs. \cite{kabuss:effectivehamiltonian, Naumann:16} it was shown that the naive detuning at the phonon frequency $\Delta=\omega_{ph}$ does not lead to maximal output intensities as the anti-Stokes resonance is shifted with respect to the pumping strength $\Omega$ and the electron-phonon coupling $g$. We show that there is an additional energy shift due to a many-particle interaction via the cavity mode resulting in a variety of resonances depending on the number of participating emitters.

\section{Collective phonon processes}\label{sec:Collective two phonon process}

We detune the optical driving frequency $\Delta$ in order to find the resonances where the cavity phonons show coherent statistics. We show these resonances in Fig. \ref{figure:spectra}. In contrast to the naive assumption, that the system is lasing at the anti-Stokes resonance ($\Delta/\omega_{ph}=1$), the frequencies are red shifted with respect to the pumping strength and the electron-phonon coupling which was already shown for the one emitter case (blue, solid) by Ref. \cite{kabuss:effectivehamiltonian}. For two emitters (red, dotted), we find a characteristic feature for the phonon laser: Instead of one single resonance, there are two resonances close to $\Delta\approx\omega_{ph}$. One resonance is exactly at the single emitter resonance with slightly higher output. Additionally, a second resonance is visible at lower driving frequencies which overlaps with the single emitter resonance. This explains the higher output compared to the single emitter case at the single emitter resonance. For three emitters (yellow, dashed) we observe the same tendency. The single as well as the two emitter resonances are apparent and additionally a third resonance at a lower driving frequency shows up. In the following, we call the resonance caused by additional emitters collective resonances. Due to the many-emitter character, these resonances are caused by a collective phonon emission of the emitters respectively. We observe a narrowing of the linewidth for these collective resonances compared to the single emitter case. \\
In order to identify the statistics of the cavity phonons at the respective resonances, we show in Fig. \ref{figure:spectra}(bottom) the autocorrelation function $g^2(0)$ \cite{2017arXiv170603777H, 2017arXiv170400446D}. The steady state value is defined as
\begin{align}
g^2(\tau)=\lim_{t \to \infty}\frac{\left\langle b^\dagger(t)b^\dagger(t+\tau)b(t+\tau)b(t)\right\rangle}{\left\langle b^\dagger(t)b(t)\right\rangle^2}~.
\end{align}
We consider only the case $\tau=0$. In Fig. \ref{figure:spectra}, we observe always $g^2(0)\ge 1$. A value $g^2(0)= 1$ is identified with coherent statistics. We observe $g^2(0)= 1$ at the resonances, where the phonon number shows a peak for the collective resonances as well as for the single emitter resonance. A value $g^2(0)> 1$ is identified with thermal statistics which is always the case if no resonance is addressed. \\
For higher driving frequencies close to $\Delta\approx 2\omega_{ph}$ the same pattern of resonances appear with coherent phonon statistics $g^2(0)= 1$. In contrast to the one phonon resonance, the linewidth is narrowing at the two phonon resonance but the number of phonons is twice as high. For the three emitter case we could not find a third resonance for these pumping strength. At higher pumping strengths, also the third resonance appears.\\
These findings prove that lasing at the two phonon resonance is theoretically possible. Furthermore, we find an interesting many-emitter effect, namely the different resonances, which we investigate in the following.\\
\subsection{Effective Hamiltonian approach}
In order to explain the different resonances, we derive an effective Hamiltonian with respect to the excitation conditon $\Delta\approx\omega_{ph}$, similar to Ref. \cite{kabuss:effectivehamiltonian}. In contrast to the full Hamiltonian in Eqs. \eqref{eq:H_0} and \eqref{eq:H_I} which includes all processes, we restrict the Hamiltonian to the second order process of cavity phonon generation in order to achieve a Jaynes-Cummings like interaction. We transform $e^{iS}\left(\mathcal{H}_0+\mathcal{H}_I\right)e^{-iS}=\mathcal{H}^{eff}$ with \emph{S} chosen to eliminate the first order electronic processes and cut the expansion after the second order. Therefore, the conditions
\begin{align}
\left[iS,\mathcal{H}_0\right]&= - \mathcal{H}_I~, 
\label{eq:first_condition}
\end{align}
and
\begin{align}
\left[iS\left[iS,\mathcal{H}_0\right]\right]&=-\left[iS,\mathcal{H}_I\right]~,
\label{eq:second_condition}
\end{align}  
determine the transformation operator
\begin{align}
S=\sum_{i=1}^{N_{QD}}\left(-\frac{i\Omega}{\Delta}\sigma^i_-+\frac{i\Omega}{\Delta}\sigma^{i}_+-\frac{ig}{\omega_{ph}}\sigma^{i}_+\sigma^i_- b^\dagger+\frac{ig}{\omega_{ph}}\sigma^{i}_+\sigma^i_- b~\right).
\end{align}
Applying the conditions \eqref{eq:first_condition} and \eqref{eq:second_condition}, the effective Hamiltonian is derived via 
\begin{align}
\mathcal{H}^{eff}=\mathcal{H}_0+\frac{1}{2}\left[iS, \mathcal{H}_I\right]~.
\end{align}
In contrast to ref. \cite{kabuss:effectivehamiltonian} the Hamiltonian contains the sum over the number of emitters which leads to a fundamental difference: The quantum dots are not directly coupled, but indirectly via the phonon operators $b^{(\dagger)}$. Therefore, if the electronic operators  are coupled to the phonon mode, the QD-QD interaction does not vanish. As the electron-phonon coupling is diagonal, the different QDs couple only via the phonon assisted electronic densities  $\sigma^{i}_+\sigma^i_- b^{(\dagger)}$. As our focus lies here on the many-emitter effects, we split the Hamiltonian into three parts to highlight the QD-QD interaction
\begin{align}
\mathcal{H}^{eff}&=\mathcal{H}^{eff}_{0}+\mathcal{H}^{eff}_{I}+\mathcal{H}^{eff}_{QD-QD}~,
\end{align} 
with $\mathcal{H}^{eff}_{QD-QD}$ as the many particle interaction term
\begin{align}
\label{eq:h_eff0}
\mathcal{H}^{eff}_{0}=&\sum_{i=1}^{N_{QD}}\hbar \omega_{eff}\sigma^{i}_+\sigma^i_-+\hbar \omega_{ph} b^\dagger b~,\\
\label{eq:h_effww}
\mathcal{H}^{eff}_{I}=&\sum_{i=1}^{N_{QD}}\hbar
g_{eff}\left(\sigma^i_-b+\sigma^{i}_+b^\dagger\right)~,\\
\mathcal{H}^{eff}_{QD-QD}=&-\sum_{i\neq
j}^{N_{QD}}\frac{g^2}{\omega_{ph}}\left(\sigma^{i}_+\sigma^i_-\otimes
\sigma^{j}_+\sigma^j_-\right)
\label{Hamiltonianbetweenqp}~,
\end{align}
Due to the assumption of identical emitters, Eq. \eqref{eq:h_eff0} and \eqref{eq:h_effww} do not differ except for the sum from the single emitter limes of ref. \cite{kabuss:effectivehamiltonian}. $\mathcal{H}^{eff}$ now is a Tavis-Cummings like Hamiltonian with two level systems of energy gap
\begin{align}
\omega_{eff}=-\Delta-\frac{2
|\Omega|^2}{\Delta}-\frac{g^2}{\omega_{ph}}~, \label{effective_energy}
\end{align}
and effective coupling strengths
\begin{align}
g_{eff}=\frac{\Omega
g}{2}\left(\frac{1}{\Delta}+\frac{1}{\omega_{ph}}\right)~.
\end{align}
In contrast to the Tavis-Cummings model, the phonon emission and absorption is reversed due to the conduction band electron-phonon coupling: $\sigma^{i}_+$ creates a phonon, while $\sigma^{i}_-$ annihilates one with $g_{eff}$.\\
Eq. \eqref{Hamiltonianbetweenqp} describes the interaction between the quantum dots via the phonon field. This term is not negligible as the electronic operators interact with phonons. We highlight that Eq. \eqref{Hamiltonianbetweenqp} is in addition to the common Tavis-Cummings model and is a characteristic feature of the phonon laser due to the diagonal electron-phonon coupling.  We show that this density-density shift of the resonance indeed has an effect on the system. In Eq. \eqref{effective_energy} it becomes clear that the energy gap is dependent on the phonon coupling $g$ and the pumping strength $\Omega$. Additionally, the interaction between quantum dots also participates to the energy gap and has the same prefactors than the last summand in Eq. \eqref{effective_energy}.\\
\begin{figure}[!htb]
\centering
\resizebox{0.48\textwidth}{!}{\input{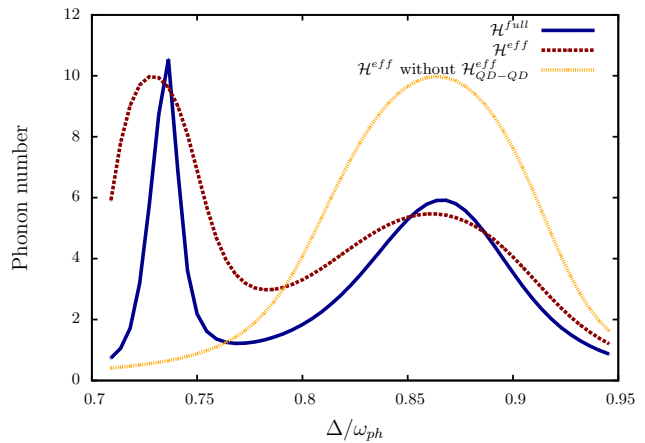}}
\caption{Validation of the effective Hamiltonian for two emitters: Comparing the full (blue, solid)  with the effective Hamiltonian (red, dotted) both models are in good agreement. The resonance for maximal phonon numbers is slightly red shifted for both resonances for the effective Hamiltonian. Also it predicts a smaller amount of phonons at the maximum while the linewidth is broadened. Turning off the QD-QD interaction, the effective Hamiltonian (yellow dotted) shows only the single emitter resonance. Note that for clarity the coupling strength $g$ is twice as large as before.} 
\label{figure:effective_hamiltonian}
\end{figure}
In Fig. \ref{figure:effective_hamiltonian} we validate the effective Hamiltonian. Now that we calculated the effective model, we can turn the QD-QD interaction of the effective Hamiltonian on and off to investigate its effect on the system response under varying the optical detuning. Considering the effective QD-QD interaction, we observe a good qualitative agreement of both models even if we have assumed the system to be at the anti-stokes resonance $\Delta=\omega_{ph}$. Neglecting the QD-QD interaction in Eq. \eqref{Hamiltonianbetweenqp} we find a different behavior. The collective resonance at the lower frequency disappears and only the single emitter resonance is predicted. In contrast to the full model, the effective model without $H^{eff}_{QD-QD}$ shows twice as much phonons at the single emitter resonance and a broadened linewidth.  At the collective resonance, the full model and the effective model with QD-QD interaction also show twice as much phonons. We conclude that the scaling with $g^2/\omega_{ph}$ with the density of both quantum dots is responsible for the collective resonance. 
\subsection{Collective resonance}
\begin{figure*}[htb!]
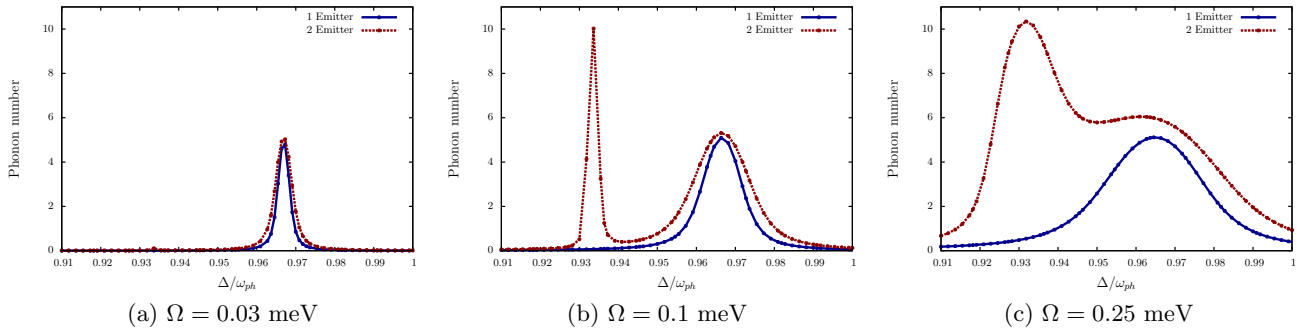

\centering
\resizebox{0.32\textwidth}{!}{\input{Phononnumber_300}}
\resizebox{0.32\textwidth}{!}{\input{Phononnumber_1000}}
\resizebox{0.32\textwidth}{!}{\input{Phononnumber_2500}}
(a) $\Omega=0.03~\text{meV}$ \hspace{3.2cm}(b) $\Omega=0.1~\text{meV}$ \hspace{3.2cm} (c) $\Omega=0.25~\text{meV}$
\caption{Pump dependency of the one (blue solid) and two emitter (red dotted) case. For low pumping strengths, only the resonance of the one emitter limes is lasing. Increasing the pumping strengths, also the collective resonance in eq. \eqref{Hamiltonianbetweenqp} is being populated and starts lasing. For high pumping strength, the collective resonance shows double the phonon number of the one Emitter case. Also the resonances are broadening due to the stronger pumping, such that both resonances overlap. } 
\label{figure:pump}
\end{figure*}
In analogy to the lasing cycle shown in Fig. \ref{fig:model-system}(b), we now focus on the many emitter case for a deeper understanding of the collective resonances. We study two cases: First, the detuning at the single emitter resonance. Either one quantum dot is excited at $|c_i, n\rangle$ and participates in a stimulated phonon emission process by interacting with a cavity phonon resulting in $|c_i, n+1\rangle$. Second, the detuning at the collective resonance. Both quantum dots are excited at $|c_1,c_2, n\rangle$ and both generate a phonon collectively via interacting with the cavity field resulting in $|c_1,c_2, n+2\rangle$. For the second case, the resonance is shifted additionally due to Eq. \eqref{Hamiltonianbetweenqp}. Because of the collective two phonon process, the phonon number is doubled in comparison to the first case. Furthermore, we observe a linewidth narrowing which indicates a longer lifetime of both QDs being in the state $|c_1,c_2, n\rangle$.\\
The many-particle interaction is dependent on the number of emitters and the occupation of the excited states of all emitters (cp. Eq. \eqref{Hamiltonianbetweenqp}). Due to coupling to the conduction band density, both emitters have to be excited for collective phonon emission. We show in Fig. \ref{figure:pump} that indeed the collective resonance disappears at low pumping strengths where for driving at the single emitter resonance, the system is already lasing. Increasing the pumping strength addresses the collective resonance as it becomes now probable for both emitters to be excited at the same time. The phonon output is increasing until it reaches twice the amount of phonons of the single emitter resonance while the single emitter resonance only increases due to the overlap with the collective resonance. Furthermore, increased pumping also results in linewidth broadening. \\
We conclude that the conditions for collective two phonon emission are high pumping strengths such that it is probable for both emitters being in the excited state $|c_1,c_2, n\rangle$ as well as a detuning at the collective resonance 
\begin{align}
\Delta^{collective}\approx\omega_{ph}-\frac{\textbf{2}g^2}{\omega_{ph}}\label{eq:collective_resonance}
\end{align} due to Eq. \eqref{Hamiltonianbetweenqp}. Therefore, driving the system at the collective resonance with $\Omega$ being strong enough results in stimulated two phonon emission of both emitters collectively. The single emitter resonance only addresses stimulated single phonon emission of either one of the emitters.\\
These findings prove that the many-emitter case also includes the single emitter resonance. We do not need to know the exact collective shift dependecies in order to generate stimulated phonon emission in the many-emitter case. One can detune at the single emitter resonance and generate the same output for a higher number of emitters. This proves the robustness of the many-emitter case against the collective interaction as the single emitter phonon emission is still included in the many-emitter case and can be addressed via the detuning. We want to mention, that the position of the resonances are highly sensible on the electron-phonon coupling $g$ as it scales with the square. In contrast to Ref. \cite{PhysRevLett.109.054301} we have chosen a higher $g$ as for smaller electron-phonon coupling the collective and the single emitter resonance move together and smear out. Therefore, a smaller $g$ also supports the robustness of the many-emitter phonon laser against these collective effects as the collective resonance is also addressed at the single emitter resonance due to the overlap. However, the electron-phonon coupling is not readily accessible in experiments. We suggest that it may be possible to deduce the electron-phonon coupling $g$ from the splitting between the resonances. For higher $g$ the resonances split up even stronger than in the presented case. If $g$ is high such that the resonances split up, one needs to consider the collective shift in Eq. \eqref{Hamiltonianbetweenqp}, using the many-emitter setup for maximal enhancement. This results in collective $N$-phonon emission of $N$-emitters and therefore in higher phonon intensities. \\  

\subsection{Two phonon resonance}
Choosing a detuning close to the single phonon resonance $\Delta\approx\omega_{ph}$ results in many resonances dependent on the number of emitters in the system. We have identified the resonances with collective N-phonon emission due to the many-emitter setup. Additionally, in Fig. \ref{figure:spectra} we see an enhancement of the phonon intensities at $\Delta \approx 2 \omega_{ph}$ which we call the two phonon resonance. We have to differentiate again between the two cases of the previous section, as other resonances are also present.\\
For the first case with detuning at the single emitter resonance we identify the process with a two phonon generation of one single emitter, where one QD is in the state $|c_i, n\rangle$ and generates two phonons via interacting with the cavity phonons resulting in $|c_i, n+2\rangle$. In Fig. \ref{figure:spectra} it is shown that driving at the two phonon resonance also results in coherent phonon statistics. Therefore, we have two different resonances resulting in a two phonon process. First, the collective two emitter two phonon generation at $\Delta\approx \omega_{ph}$ and second the single emitter two phonon generation at $\Delta \approx 2 \omega_{ph}$. Comparing both resonances in Fig. \ref{figure:spectra} we identify the same line shape if we exclude the overlap with the one phonon resonance. Both resonances result in the same amount of phonons as well as nearly the same linewidth. However, the excitation scheme is completely different. On one hand the two phonon generation results from collective stimulated two phonon emission of two emitters and on the other hand from a stimulated two phonon generation of one single emitter. Therefore, it is surprising that both resonances are that alike and result in coherent statistics. \\
For the second case, addressing the collective two phonon resonance we identify it with a four phonon process. Here, both QDs are excited at $|c_1,c_2, n\rangle$. Interacting with the cavity phonons results in a collective four phonon emission of both quantum dots $|c_1,c_2, n+4\rangle$. The number of phonons is twice as high as for the two phonon processes and four times higher than the single phonon emission process. Note, that its linewidth is very narrow in comparison to the other resonances. The resonance is very sharp and it might be difficult to address it via the optical detuning. Following this argumentation for the collective resonance for three emitters at the two phonon resonance, we identify it with a six phonon emission process. The QDs have to be in the state $|c_1,c_2,c_3, n\rangle$ for collective three emitter two phonon emission $|c_1,c_2,c_3, n+6\rangle$. For the investigated pump strength in Fig. \ref{figure:spectra} this state is not populated. Increasing the pumping strength we also observe this collective six phonon resonance. However, its linewidth is even narrower than the other resonances which makes it more difficult to address. 

\section{Non identical emitters}\label{nonidentical emitters}

In this section we investigate the robustness of the collective resonances against effects resulting from non identical emitters. Realistic QDs differ in size and have different spatial positions which result in a variety of transition frequencies and  different electron-phonon couplings $g$ \cite{Bimberg:QD, michler:QD, Jahnke:QD, doi:10.1063/1.3117236}. We do not investigate effects resulting from inhomogeneous  broadening but provide an estimation which difference in frequency or electron-phonon coupling of two individual QDs can be tolerated in order to still observe collective phonon generation.
\begin{figure}[!htb]
\centering
\includegraphics[width=0.48\textwidth]{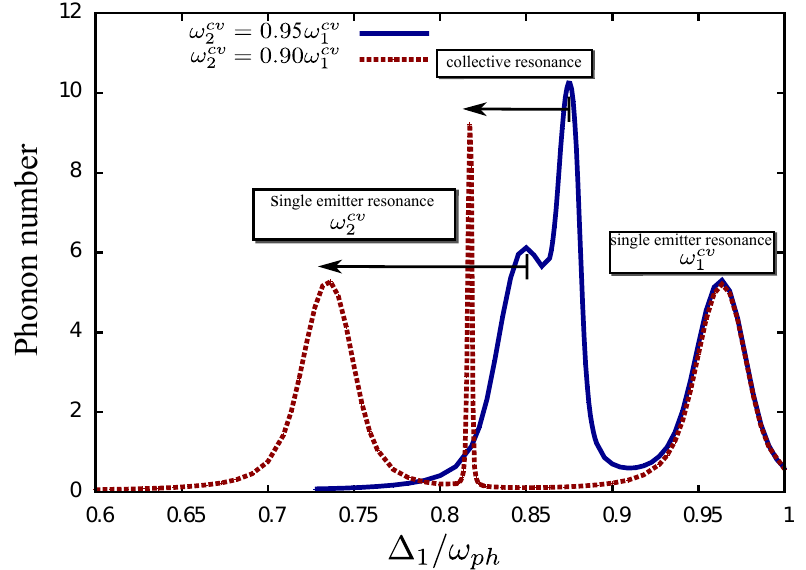}
\caption{Optical detuning versus phonon number for two emitters which differ in the two-level band gap frequency $\omega^{cv}$. We show a deviation of the second QD of 5\% (blue,solid) and 10\% (red, dotted), while the first QD is kept at the same transition frequency as before. The single emitter resonances split up due to the different transition frequencies. The collective resonance is still visible in both cases. For increasing difference of the transition frequencies, the linewidth of the collective resonance narrows and the phonon number decreases. }  
\label{figure:omcvdifference}
\end{figure}
We investigate a difference in the transition frequencies of the two QDs, where $\omega^{cv}_2$ of the second QD deviates 5\% (blue,solid) or 10\% (red, dotted) from $\omega^{cv}_1$ of the first QD which we show in Fig. \ref{figure:omcvdifference}. A difference in electronic transition frequencies results in a strong splitting of the two single emitter resonances. The single emitter resonance of the second QD is located at a lower frequency due to $\omega^{cv}_2<\omega^{cv}_1$ and even lower than the collective resonance. For collective phonon generation, both QDs participate which suggests that the location of the collective resonance is a mixture of both transition frequencies. The positions of the collective resonances can be estimated well for both cases in Fig. \ref{figure:omcvdifference} with
\begin{align}
\Delta^{collective}=\omega_{ph}-\frac{\omega^{cv}_1-\omega^{cv}_2}{2}-\frac{2g^2}{\omega_{ph}}~.
\end{align}
Note that we have taken $\omega^{cv}_1$ as reference within the figures ($\Delta_1=\omega_L-\omega^{cv}_1$). For equal QDs, the second summand vanishes and yields the resonance determined by Eq. \eqref{eq:collective_resonance}. As indicated by the narrowing of the collective resonance in Fig. \ref{figure:omcvdifference} for 10\% difference (red, dotted), we expect that for greater difference in the transition frequencies, the collective resonance narrows more and the phonon number decreases until it vanishes.\\
\textit{In contrast to the frequency difference, collective phonon generation is rather robust against difference in the electron-phonon coupling.}
\begin{figure}[!htb]
\centering
\includegraphics[width=0.48\textwidth]{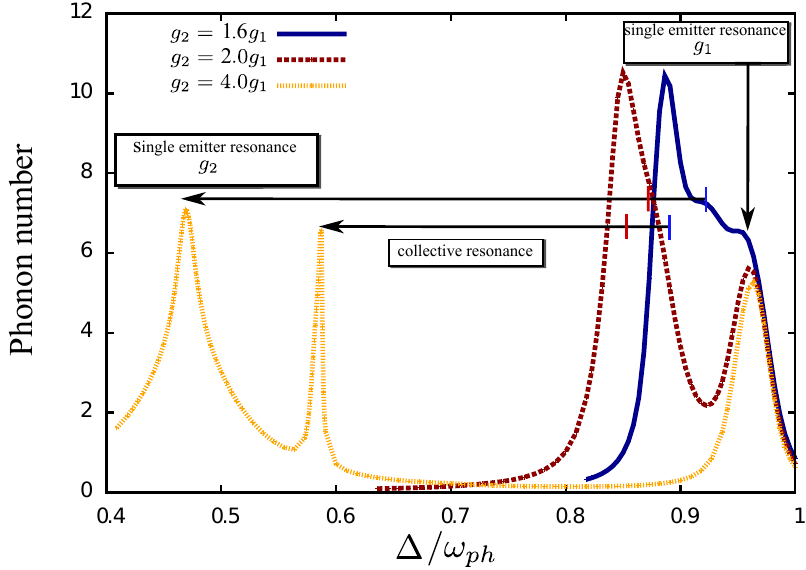}
\caption{Optical detuning versus phonon number for two emitters which differ in the electron-phonon coupling $g$. For small differences $g_2=1.6g_1$ (blue, solid) the single emitter resonances split up but still overlap with the collective resonance. For $g_2=2g_1$ (red, dotted) the single emitter resonance of $g_2$ is overlapping with the collective resonance. For $g_2=4g_1$ (yellow, dashed), all resonances split up and the single emitter resonance is at lower $\Delta$ than the collective resonance. Furthermore, the output of the collective resonance is smaller than the single emitter resonance of $g_2$. }  
\label{figure:gdifference}
\end{figure}
In Fig. \ref{figure:gdifference}, we show the behavior if the electron-phonon coupling $g$ of one QD is different while the other one is kept at the same $g$ as before (transition frequencies are equal). We show three cases:  $g_2=1.6g_1$ (blue, solid), $g_2=2g_1$ (red, dotted) and $g_2=4g_1$ (yellow, dashed) in order to estimate the position of the collective resonances. The single emitter resonances split up due to the different $g_i$. This can be estimated from the respective effective single emitter Hamiltonian in Eq. \eqref{effective_energy} (excluding the comparably small shift due to the pump $\Omega$)
\begin{align}
\Delta_i^{single}\approx\omega_{ph}-\frac{g_i^2}{\omega_{ph}}~,\label{eq:single_resonance_non-identical}
\end{align}
as collective processes do not play a role.
However, for the collective resonances, the effective Hamiltonian in Eqs. \eqref{eq:h_eff0}-\eqref{Hamiltonianbetweenqp} is only valid for identical emitters. Therefore, the collective resonance cannot be determined by Eq. \eqref{eq:collective_resonance}.
Altogether, this results for the two emitter case in three resonances: The two single emitter resonances due to different $g_i$ and the collective resonance. For a difference $g_2=1.6g_1$ (blue, solid), the three resonances are overlapping, but in contrast to Fig. \ref{figure:spectra}(red, dotted line), there is clearly an additional resonance. The single emitter resonance of the first QD is at the same position as before, but with higher output due to the overlap with the single emitter resonance of the second QD. This resonance is located at lower frequencies than for the first QD but with higher output, due to the higher $g_2$. The collective resonance shows the highest output. For a greater difference between $g_1$ and $g_2$, the position of the peaks split up more. For $g_2=2g_1$ (red, dotted), the single emitter resonance of the second QD is very close to the collective resonance at a lower frequency. An interesting effect appears in case of $g_2=4g_1$ (yellow, dashed): The single emitter resonance of the second QD is located at a smaller frequency than the collective resonance. Also the phonon number of the single emitter resonance of the second QD is now higher than the collective resonance. The reason for the higher single emitter resonance is $g_2>g_1$. One might expect that a higher $g_2$ would result as well in a higher collective resonance. We suppose, as it can be seen in Fig. \ref{figure:gdifference}, that a great difference between the QDs for the electron phonon coupling $g$ results in a narrower collective resonance with lesser output as it was already the case for increasing difference in the transition frequencies (Fig. \ref{figure:omcvdifference}). For all cases shown in Fig. \ref{figure:gdifference}  we estimate the position of the collective resonance by 
\begin{align}
\Delta^{collective}\approx\omega_{ph}-\frac{\left(g_1+g_2\right)^2}{2\omega_{ph}}~.\label{eq:collective_resonance_non-identical}
\end{align}
Note that in case of identical emitters this agrees with the collective resonance determined  before in Eq. \eqref{eq:collective_resonance}.
 We conclude, that non identical QDs do not destroy the effect of collective phonon generation. Although a great difference in the QD transition frequencies ($>10\%$) narrows the linewidth of the collective resonance remarkably. The collective resonance is more robust against a difference in the electron-phonon coupling. A difference up to $\sim400\%$ can be tolerated before the resonance is narrowing. 
\section{Quantum yield}\label{sec:Collective quantum yield enhancement}

We have identified the different resonances close to $\Delta\approx\omega_{ph}$ in Fig. \ref{figure:spectra} with collective phonon emission. If the emitters are all in the excited state $|c_i,... ,c_N, n\rangle$, they collectively generate stimulated phonon emission at the respective resonance due to Eq. \eqref{Hamiltonianbetweenqp}. In order to unravel the enhancement of the many emitter setup, we investigate the quantum yield. We compare driving of the single emitter and the collective resonance of the two emitter case with driving at the single emitter resonance of the one emitter case. In Fig. \ref{figure:pump} it was shown that the collective resonance is lasing only at high pump powers. Therefore, we are interested in the quantum yield depending on the external pump power. In order to compare the resulting phonon numbers, we define a phonance witness in analogy to the radiance witness of Ref. \cite{2017arXiv170205392P}
\begin{align}
R=\frac{\langle b^\dagger b\rangle_2-2\langle b^\dagger b\rangle_1}{2\langle b^\dagger b\rangle_1}~, \label{witness}
\end{align}
with $\langle b^\dagger b\rangle_2$ being the mean phonon number of the two emitter setup and $\langle b^\dagger b\rangle_1$ for one emitter. Thus, $\langle b^\dagger b\rangle_2$ yields the correlated phonon number, whereas $2\langle b^\dagger b\rangle_1$ represents the expected uncorrelated phonon number for two emitters. Due to the normalization with the uncorrelated phonon number, $R=1$ signifies a collective enhanced scaling of the phonon number (with $\langle b^\dagger b\rangle_2=4\langle b^\dagger b\rangle_1$). The uncorrelated case $R=0$ is related to the expected lasing behavior with linear increase of the phonon number with the number of emitters $\langle b^\dagger b\rangle_N\sim N$. Any $R<0$ is a subradiant like behavior, where the number of emitters suppresses the output \cite{2017arXiv170502889G}. \\
\begin{figure}[!htb]
\centering
\resizebox{0.48\textwidth}{!}{\input{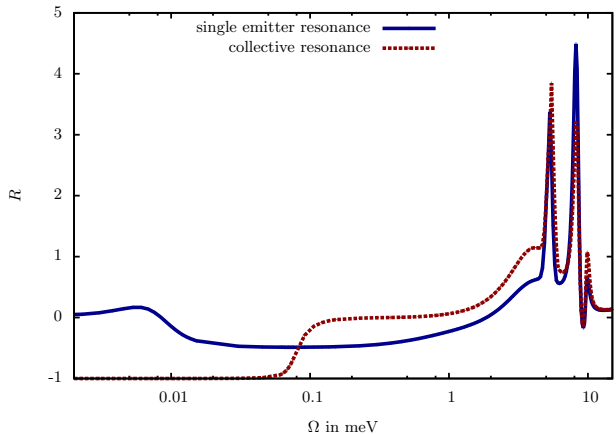}}
\caption{Phonance witness  $R$ of the two emitter phonon laser versus the external optical lasing pump power $\Omega$. We compare driving at the single emitter resonance (blue) with driving at the collective resonance (red dashed). Note that the one emitter case $\langle b^\dagger b\rangle_1$ is always driven at the single emitter resonance. After $\Omega= 0.1$ meV the collective resonance shows a higher quantum yield $R \ge 0$. After $\Omega= 2$ meV both cases show collective enhancements with $R>0$, including two peaks with even $R>1$.}  
\label{figure:wittness}
\end{figure}
The phonance witness is shown in Fig. \ref{figure:wittness}. First, we consider the phonance witness with a detuning at the single emitter resonance $\Delta=\omega_{ph}-\frac{g^2}{\omega_{ph}}$ for one and two emitters (blue, solid). For low pump powers, we observe $R >0$ what we explain with a different lasing threshold. Due to the two emitter correlation, it enters the lasing regime at lower pump powers in comparison to the single emitter setup and thus showing a higher phonon number. Increasing $\Omega$, also the one emitter setup enters the lasing regime and $R$ drops below zero and shows subradiant like behavior. The reason for $R <0$ is the driving at the single emitter resonance where either one of the emitters emits a phonon with effectively being a single emitter setup as explained in the previous section. Therefore, the phonon number of the two emitter setup is equal to the one emitter case with no many-emitter enhancement. With increasing pump power $\Omega>0.2$ meV, the peaks of the single emitter and the collective resonance are overlapping what can be seen as well in Fig. \ref{figure:pump}. Due to the increased pump and the peak overlap, the phonon number at the single emitter resonance for two emitters is higher than for one emitter and as a consequence $R$ increases.  At $\Omega\approx 2$ meV, the one emitter setup stops lasing due to the self quenching behavior \cite{PhysRevLett.109.054301, kabuss:effectivehamiltonian, Naumann:16} while the two emitter setup is still lasing and the phonance witness increases as it was the case for very low pump powers. Shortly before the two emitter case stops lasing as well, the phonance witness shows a peak and jumps to a value $R>1$ indicating a high collective enhancement. High pump powers also address the two phonon resonance, as already shown by Ref. \cite{Naumann:16}. The first peak is associated with the collective two phonon resonance which is only present for $N\ge 2$. Thus, two emitters are lasing at the collective two phonon resonance while one emitter is within the self quenching regime without lasing action which is the reason for the high phonance witness. For higher pump powers, the witness decreases again before reaching a second and a third peak. Both peaks result from the two phonon single emitter resonance which is addressed by the high pump. The reason for the two peaks is the different linewidth (with pump power at the x-axis) for one and two emitters. For one emitter, the two phonon resonance shows a narrower peak, thus resulting in two peaks in the phonance wittness. For higher pump powers the resonance is shifting out of the two phonon resonance and both two and one emitter are within the self quenching regime again with $R\approx 0$. Note, that Ref. \cite{Naumann:16} is based on a coherent factorization and on the assumption of identical behavior of the QDs. Thus, observing always lasing at the collective resonance for higher number of emitters. In contrast, we unravel the different shift dependencies of the many emitter case via the full quantum mechanical treatment and also prove coherent statistics at the two phonon resonance (Fig. \ref{figure:spectra}). \\
Now let us consider the phonance witness with driving at the collective resonance $\Delta=\omega_{ph}-\frac{2g^2}{\omega_{ph}}$ in Fig. \ref{figure:wittness} (red, dotted).  For low pump powers, it is not probable for both emitters to be in the state $|c_1, c_2, n\rangle$ and no lasing occurs as can be seen in Fig. \ref{figure:pump}. Therefore, $R=-1$ as the one emitter setup driven at the single emitter resonance shows phonon lasing. At $\Omega\approx0.05$ meV, a detuning at the collective resonance also starts lasing as it is now probable for both emitters being in the state $|c_1, c_2, n\rangle$ (cp. Eq. \eqref{Hamiltonianbetweenqp}). $R$ increases until the two emitter setup reaches twice the phonon number at $\Omega\approx 0.2$ and thus $R=0$ as for expected many-emitter lasing. This clarifies the importance of driving at the collective resonance for the usual many emitter enhancement $\sim N$ of lasing action. From this point on, increasing $\Omega$ results in a phonance witness similar to driving at the single emitter resonance. The phonon number of the single emitter setup decreases at lower pump powers and therefore $R$ increases for both cases. With one difference: Driving at the collective resonance results in higher output, therefore also $R$ is higher than for driving at the single emitter resonance and $R >1$ is reached before the two phonon resonance is addressed. Note that driving at the collective resonance reaches the two phonon resonance at slightly higher pump powers, because the lasing frequency is already detuned with respect to the collective resonance. Besides that, both, driving at the single as well as at the collective resonance shows nearly the same phonance witness  at the two phonon resonance addressed via the pump. Therefore, for high pump powers it becomes nearly independent of the optical detuning. This is a strong statement and proves again the robustness of the many emitter phonon laser. Without careful adjustment of the detuning, increasing the pump such that the two phonon resonance is addressed results in perfect many emitter enhancement with $R>1$. We assume that this behavior is also the same for higher numbers of emitters as it becomes independent of the detuning. For best enhancement at lower pump powers, the number of emitters have to be considered in the optical detuning. Once the lasing at the two phonon resonance is reached via increased pump, the adjustment of the detuning becomes redundant as perfect enhancement can be achieved via strong optical pumping at the collective two phonon resonance. However, for maximal cavity intensities, the optical detuning should be adjusted as well.

\section{Conclusion}
In conclusion, we considered the few emitter regime of phonon lasing. In contrast to the single emitter phonon laser, additional resonances appear, dependent on the number of emitters. An effective Hamiltonian approach results in a Tavis-Cummings like interaction with an additional energy shift due to a many-emitter interaction. That clarifies that the additional resonances result from collective many-emitter phonon emission. The same pattern of resonances appears close to the two phonon resonance which results in lasing as well. In case of non identical emitters, the two single emitter resonances split up for either a difference in the electronic transition frequencies or in the electron-phonon coupling. The collective emission for two quantum dots is robust against different transition frequencies up until $\sim 10\%$ while for the electron phonon coupling differences up to $\sim400\%$ can be tolerated. Investigating the quantum yield, driving at the collective resonances results in the expected linear scaling of the phonon number with the number of emitters. A collective enhanced scaling can be achieved close to the lasing threshold, as the many-emitter laser has lower threshold. For high pump powers, the two phonon resonances are addressed and result in perfect many emitter enhancement.  
\section*{Acknowledgments}
We thank Andreas Knorr and Michael Gegg for fruitful discussions. We gratefully acknowledge the support of the
Deutsche Forschungsgemeinschaft (DFG) through the project B1 of the SFB 910 and by the school of nanophotonics (SFB 787).
\bibliography{Hauptdatei} 

%merlin.mbs apsrev4-1.bst 2010-07-25 4.21a (PWD, AO, DPC) hacked
%Control: key (0)
%Control: author (8) initials jnrlst
%Control: editor formatted (1) identically to author
%Control: production of article title (-1) disabled
%Control: page (0) single
%Control: year (1) truncated
%Control: production of eprint (0) enabled
\begin{thebibliography}{41}%
\makeatletter
\providecommand \@ifxundefined [1]{%
 \@ifx{#1\undefined}
}%
\providecommand \@ifnum [1]{%
 \ifnum #1\expandafter \@firstoftwo
 \else \expandafter \@secondoftwo
 \fi
}%
\providecommand \@ifx [1]{%
 \ifx #1\expandafter \@firstoftwo
 \else \expandafter \@secondoftwo
 \fi
}%
\providecommand \natexlab [1]{#1}%
\providecommand \enquote  [1]{``#1''}%
\providecommand \bibnamefont  [1]{#1}%
\providecommand \bibfnamefont [1]{#1}%
\providecommand \citenamefont [1]{#1}%
\providecommand \href@noop [0]{\@secondoftwo}%
\providecommand \href [0]{\begingroup \@sanitize@url \@href}%
\providecommand \@href[1]{\@@startlink{#1}\@@href}%
\providecommand \@@href[1]{\endgroup#1\@@endlink}%
\providecommand \@sanitize@url [0]{\catcode `\\12\catcode `\$12\catcode
  `\&12\catcode `\#12\catcode `\^12\catcode `\_12\catcode `\%12\relax}%
\providecommand \@@startlink[1]{}%
\providecommand \@@endlink[0]{}%
\providecommand \url  [0]{\begingroup\@sanitize@url \@url }%
\providecommand \@url [1]{\endgroup\@href {#1}{\urlprefix }}%
\providecommand \urlprefix  [0]{URL }%
\providecommand \Eprint [0]{\href }%
\providecommand \doibase [0]{http://dx.doi.org/}%
\providecommand \selectlanguage [0]{\@gobble}%
\providecommand \bibinfo  [0]{\@secondoftwo}%
\providecommand \bibfield  [0]{\@secondoftwo}%
\providecommand \translation [1]{[#1]}%
\providecommand \BibitemOpen [0]{}%
\providecommand \bibitemStop [0]{}%
\providecommand \bibitemNoStop [0]{.\EOS\space}%
\providecommand \EOS [0]{\spacefactor3000\relax}%
\providecommand \BibitemShut  [1]{\csname bibitem#1\endcsname}%
\let\auto@bib@innerbib\@empty
%</preamble>
\bibitem [{\citenamefont {Haken}(1994)}]{Haken:Laser}%
  \BibitemOpen
  \bibfield  {author} {\bibinfo {author} {\bibfnamefont {H.}~\bibnamefont
  {Haken}},\ }\href@noop {} {\emph {\bibinfo {title} {Licht und Materie II}}}\
  (\bibinfo  {publisher} {B. I. Wissenschaftsverlag, Mannheim},\ \bibinfo
  {year} {1994})\BibitemShut {NoStop}%
\bibitem [{\citenamefont {Chow}\ \emph {et~al.}(1994)\citenamefont {Chow},
  \citenamefont {Koch},\ and\ \citenamefont {Sargent}}]{weng:laser}%
  \BibitemOpen
  \bibfield  {author} {\bibinfo {author} {\bibfnamefont {W.~W.}\ \bibnamefont
  {Chow}}, \bibinfo {author} {\bibfnamefont {S.~W.}\ \bibnamefont {Koch}}, \
  and\ \bibinfo {author} {\bibfnamefont {M.~I.}\ \bibnamefont {Sargent}},\
  }\href@noop {} {\emph {\bibinfo {title} {Semiconductor-Laser Physics}}}\
  (\bibinfo  {publisher} {Springer-Verlag Berlin Heidelberg},\ \bibinfo {year}
  {1994})\BibitemShut {NoStop}%
\bibitem [{\citenamefont {Khurgin}(2010)}]{viewpoint}%
  \BibitemOpen
  \bibfield  {author} {\bibinfo {author} {\bibfnamefont {J.~B.}\ \bibnamefont
  {Khurgin}},\ }\href@noop {} {\bibfield  {journal} {\bibinfo  {journal}
  {Physics}\ }\textbf {\bibinfo {volume} {3}},\ \bibinfo {pages} {16} (\bibinfo
  {year} {2010})}\BibitemShut {NoStop}%
\bibitem [{\citenamefont {{Vahala K.}}\ \emph {et~al.}(2009)\citenamefont
  {{Vahala K.}}, \citenamefont {{Herrmann M.}}, \citenamefont {{Knunz S.}},
  \citenamefont {{Batteiger V.}}, \citenamefont {{Saathoff G.}}, \citenamefont
  {{Hansch T. W.}},\ and\ \citenamefont {{Udem Th.}}}]{10.1038/nphys1367}%
  \BibitemOpen
  \bibfield  {author} {\bibinfo {author} {\bibnamefont {{Vahala K.}}}, \bibinfo
  {author} {\bibnamefont {{Herrmann M.}}}, \bibinfo {author} {\bibnamefont
  {{Knunz S.}}}, \bibinfo {author} {\bibnamefont {{Batteiger V.}}}, \bibinfo
  {author} {\bibnamefont {{Saathoff G.}}}, \bibinfo {author} {\bibnamefont
  {{Hansch T. W.}}}, \ and\ \bibinfo {author} {\bibnamefont {{Udem Th.}}},\
  }\href {\doibase http://dx.doi.org/10.1038/nphys1367} {\bibfield  {journal}
  {\bibinfo  {journal} {Nat Phys}\ }\textbf {\bibinfo {volume} {5}},\ \bibinfo
  {pages} {682} (\bibinfo {year} {2009})}\BibitemShut {NoStop}%
\bibitem [{\citenamefont {Mendonça}\ \emph {et~al.}(2010)\citenamefont
  {Mendonça}, \citenamefont {Terças}, \citenamefont {Brodin},\ and\
  \citenamefont {Marklund}}]{0295-5075-91-3-33001}%
  \BibitemOpen
  \bibfield  {author} {\bibinfo {author} {\bibfnamefont {J.~T.}\ \bibnamefont
  {Mendonça}}, \bibinfo {author} {\bibfnamefont {H.}~\bibnamefont {Terças}},
  \bibinfo {author} {\bibfnamefont {G.}~\bibnamefont {Brodin}}, \ and\ \bibinfo
  {author} {\bibfnamefont {M.}~\bibnamefont {Marklund}},\ }\href
  {http://stacks.iop.org/0295-5075/91/i=3/a=33001} {\bibfield  {journal}
  {\bibinfo  {journal} {EPL (Europhysics Letters)}\ }\textbf {\bibinfo {volume}
  {91}},\ \bibinfo {pages} {33001} (\bibinfo {year} {2010})}\BibitemShut
  {NoStop}%
\bibitem [{\citenamefont {Grudinin}\ \emph {et~al.}(2010)\citenamefont
  {Grudinin}, \citenamefont {Lee}, \citenamefont {Painter},\ and\ \citenamefont
  {Vahala}}]{PhysRevLett.104.083901}%
  \BibitemOpen
  \bibfield  {author} {\bibinfo {author} {\bibfnamefont {I.~S.}\ \bibnamefont
  {Grudinin}}, \bibinfo {author} {\bibfnamefont {H.}~\bibnamefont {Lee}},
  \bibinfo {author} {\bibfnamefont {O.}~\bibnamefont {Painter}}, \ and\
  \bibinfo {author} {\bibfnamefont {K.~J.}\ \bibnamefont {Vahala}},\ }\href
  {\doibase 10.1103/PhysRevLett.104.083901} {\bibfield  {journal} {\bibinfo
  {journal} {Phys. Rev. Lett.}\ }\textbf {\bibinfo {volume} {104}},\ \bibinfo
  {pages} {083901} (\bibinfo {year} {2010})}\BibitemShut {NoStop}%
\bibitem [{\citenamefont {Kepesidis}\ \emph {et~al.}(2013)\citenamefont
  {Kepesidis}, \citenamefont {Bennett}, \citenamefont {Portolan}, \citenamefont
  {Lukin},\ and\ \citenamefont {Rabl}}]{PhysRevB.88.064105}%
  \BibitemOpen
  \bibfield  {author} {\bibinfo {author} {\bibfnamefont {K.~V.}\ \bibnamefont
  {Kepesidis}}, \bibinfo {author} {\bibfnamefont {S.~D.}\ \bibnamefont
  {Bennett}}, \bibinfo {author} {\bibfnamefont {S.}~\bibnamefont {Portolan}},
  \bibinfo {author} {\bibfnamefont {M.~D.}\ \bibnamefont {Lukin}}, \ and\
  \bibinfo {author} {\bibfnamefont {P.}~\bibnamefont {Rabl}},\ }\href {\doibase
  10.1103/PhysRevB.88.064105} {\bibfield  {journal} {\bibinfo  {journal} {Phys.
  Rev. B}\ }\textbf {\bibinfo {volume} {88}},\ \bibinfo {pages} {064105}
  (\bibinfo {year} {2013})}\BibitemShut {NoStop}%
\bibitem [{\citenamefont {Mahboob}\ \emph {et~al.}(2013)\citenamefont
  {Mahboob}, \citenamefont {Nishiguchi}, \citenamefont {Fujiwara},\ and\
  \citenamefont {Yamaguchi}}]{PhysRevLett.110.127202}%
  \BibitemOpen
  \bibfield  {author} {\bibinfo {author} {\bibfnamefont {I.}~\bibnamefont
  {Mahboob}}, \bibinfo {author} {\bibfnamefont {K.}~\bibnamefont {Nishiguchi}},
  \bibinfo {author} {\bibfnamefont {A.}~\bibnamefont {Fujiwara}}, \ and\
  \bibinfo {author} {\bibfnamefont {H.}~\bibnamefont {Yamaguchi}},\ }\href
  {\doibase 10.1103/PhysRevLett.110.127202} {\bibfield  {journal} {\bibinfo
  {journal} {Phys. Rev. Lett.}\ }\textbf {\bibinfo {volume} {110}},\ \bibinfo
  {pages} {127202} (\bibinfo {year} {2013})}\BibitemShut {NoStop}%
\bibitem [{\citenamefont {Camps}\ \emph {et~al.}(2001)\citenamefont {Camps},
  \citenamefont {Makler}, \citenamefont {Pastawski},\ and\ \citenamefont
  {Foa~Torres}}]{PhysRevB.64.125311}%
  \BibitemOpen
  \bibfield  {author} {\bibinfo {author} {\bibfnamefont {I.}~\bibnamefont
  {Camps}}, \bibinfo {author} {\bibfnamefont {S.~S.}\ \bibnamefont {Makler}},
  \bibinfo {author} {\bibfnamefont {H.~M.}\ \bibnamefont {Pastawski}}, \ and\
  \bibinfo {author} {\bibfnamefont {L.~E.~F.}\ \bibnamefont {Foa~Torres}},\
  }\href {\doibase 10.1103/PhysRevB.64.125311} {\bibfield  {journal} {\bibinfo
  {journal} {Phys. Rev. B}\ }\textbf {\bibinfo {volume} {64}},\ \bibinfo
  {pages} {125311} (\bibinfo {year} {2001})}\BibitemShut {NoStop}%
\bibitem [{\citenamefont {Liu}\ \emph {et~al.}(2003)\citenamefont {Liu},
  \citenamefont {Song}, \citenamefont {Wasilewski}, \citenamefont
  {SpringThorpe}, \citenamefont {Cao}, \citenamefont {Dharma-wardana},
  \citenamefont {Aers}, \citenamefont {Lockwood},\ and\ \citenamefont
  {Gupta}}]{PhysRevLett.90.077402}%
  \BibitemOpen
  \bibfield  {author} {\bibinfo {author} {\bibfnamefont {H.~C.}\ \bibnamefont
  {Liu}}, \bibinfo {author} {\bibfnamefont {C.~Y.}\ \bibnamefont {Song}},
  \bibinfo {author} {\bibfnamefont {Z.~R.}\ \bibnamefont {Wasilewski}},
  \bibinfo {author} {\bibfnamefont {A.~J.}\ \bibnamefont {SpringThorpe}},
  \bibinfo {author} {\bibfnamefont {J.~C.}\ \bibnamefont {Cao}}, \bibinfo
  {author} {\bibfnamefont {C.}~\bibnamefont {Dharma-wardana}}, \bibinfo
  {author} {\bibfnamefont {G.~C.}\ \bibnamefont {Aers}}, \bibinfo {author}
  {\bibfnamefont {D.~J.}\ \bibnamefont {Lockwood}}, \ and\ \bibinfo {author}
  {\bibfnamefont {J.~A.}\ \bibnamefont {Gupta}},\ }\href {\doibase
  10.1103/PhysRevLett.90.077402} {\bibfield  {journal} {\bibinfo  {journal}
  {Phys. Rev. Lett.}\ }\textbf {\bibinfo {volume} {90}},\ \bibinfo {pages}
  {077402} (\bibinfo {year} {2003})}\BibitemShut {NoStop}%
\bibitem [{\citenamefont {Beardsley}\ \emph {et~al.}(2010)\citenamefont
  {Beardsley}, \citenamefont {Akimov}, \citenamefont {Henini},\ and\
  \citenamefont {Kent}}]{PhysRevLett.104.085501}%
  \BibitemOpen
  \bibfield  {author} {\bibinfo {author} {\bibfnamefont {R.~P.}\ \bibnamefont
  {Beardsley}}, \bibinfo {author} {\bibfnamefont {A.~V.}\ \bibnamefont
  {Akimov}}, \bibinfo {author} {\bibfnamefont {M.}~\bibnamefont {Henini}}, \
  and\ \bibinfo {author} {\bibfnamefont {A.~J.}\ \bibnamefont {Kent}},\ }\href
  {\doibase 10.1103/PhysRevLett.104.085501} {\bibfield  {journal} {\bibinfo
  {journal} {Phys. Rev. Lett.}\ }\textbf {\bibinfo {volume} {104}},\ \bibinfo
  {pages} {085501} (\bibinfo {year} {2010})}\BibitemShut {NoStop}%
\bibitem [{\citenamefont {Kabuss}\ \emph {et~al.}(2012)\citenamefont {Kabuss},
  \citenamefont {Carmele}, \citenamefont {Brandes},\ and\ \citenamefont
  {Knorr}}]{PhysRevLett.109.054301}%
  \BibitemOpen
  \bibfield  {author} {\bibinfo {author} {\bibfnamefont {J.}~\bibnamefont
  {Kabuss}}, \bibinfo {author} {\bibfnamefont {A.}~\bibnamefont {Carmele}},
  \bibinfo {author} {\bibfnamefont {T.}~\bibnamefont {Brandes}}, \ and\
  \bibinfo {author} {\bibfnamefont {A.}~\bibnamefont {Knorr}},\ }\href
  {\doibase 10.1103/PhysRevLett.109.054301} {\bibfield  {journal} {\bibinfo
  {journal} {Phys. Rev. Lett.}\ }\textbf {\bibinfo {volume} {109}},\ \bibinfo
  {pages} {054301} (\bibinfo {year} {2012})}\BibitemShut {NoStop}%
\bibitem [{\citenamefont {Kabuss}\ \emph {et~al.}(2013)\citenamefont {Kabuss},
  \citenamefont {Carmele},\ and\ \citenamefont
  {Knorr}}]{kabuss:effectivehamiltonian}%
  \BibitemOpen
  \bibfield  {author} {\bibinfo {author} {\bibfnamefont {J.}~\bibnamefont
  {Kabuss}}, \bibinfo {author} {\bibfnamefont {A.}~\bibnamefont {Carmele}}, \
  and\ \bibinfo {author} {\bibfnamefont {A.}~\bibnamefont {Knorr}},\ }\href
  {\doibase 10.1103/PhysRevB.88.064305} {\bibfield  {journal} {\bibinfo
  {journal} {Phys. Rev. B}\ }\textbf {\bibinfo {volume} {88}},\ \bibinfo
  {pages} {064305} (\bibinfo {year} {2013})}\BibitemShut {NoStop}%
\bibitem [{\citenamefont {Naumann}\ \emph {et~al.}(2016)\citenamefont
  {Naumann}, \citenamefont {Droenner}, \citenamefont {Chow}, \citenamefont
  {Kabuss},\ and\ \citenamefont {Carmele}}]{Naumann:16}%
  \BibitemOpen
  \bibfield  {author} {\bibinfo {author} {\bibfnamefont {N.~L.}\ \bibnamefont
  {Naumann}}, \bibinfo {author} {\bibfnamefont {L.}~\bibnamefont {Droenner}},
  \bibinfo {author} {\bibfnamefont {W.~W.}\ \bibnamefont {Chow}}, \bibinfo
  {author} {\bibfnamefont {J.}~\bibnamefont {Kabuss}}, \ and\ \bibinfo {author}
  {\bibfnamefont {A.}~\bibnamefont {Carmele}},\ }\href {\doibase
  10.1364/JOSAB.33.001492} {\bibfield  {journal} {\bibinfo  {journal} {J. Opt.
  Soc. Am. B}\ }\textbf {\bibinfo {volume} {33}},\ \bibinfo {pages} {1492}
  (\bibinfo {year} {2016})}\BibitemShut {NoStop}%
\bibitem [{\citenamefont {Trigo}\ \emph {et~al.}(2002)\citenamefont {Trigo},
  \citenamefont {Bruchhausen}, \citenamefont {Fainstein}, \citenamefont
  {Jusserand},\ and\ \citenamefont {Thierry-Mieg}}]{PhysRevLett.89.227402}%
  \BibitemOpen
  \bibfield  {author} {\bibinfo {author} {\bibfnamefont {M.}~\bibnamefont
  {Trigo}}, \bibinfo {author} {\bibfnamefont {A.}~\bibnamefont {Bruchhausen}},
  \bibinfo {author} {\bibfnamefont {A.}~\bibnamefont {Fainstein}}, \bibinfo
  {author} {\bibfnamefont {B.}~\bibnamefont {Jusserand}}, \ and\ \bibinfo
  {author} {\bibfnamefont {V.}~\bibnamefont {Thierry-Mieg}},\ }\href {\doibase
  10.1103/PhysRevLett.89.227402} {\bibfield  {journal} {\bibinfo  {journal}
  {Phys. Rev. Lett.}\ }\textbf {\bibinfo {volume} {89}},\ \bibinfo {pages}
  {227402} (\bibinfo {year} {2002})}\BibitemShut {NoStop}%
\bibitem [{\citenamefont {Lanzillotti-Kimura}\ \emph
  {et~al.}(2007)\citenamefont {Lanzillotti-Kimura}, \citenamefont {Fainstein},
  \citenamefont {Balseiro},\ and\ \citenamefont
  {Jusserand}}]{PhysRevB.75.024301}%
  \BibitemOpen
  \bibfield  {author} {\bibinfo {author} {\bibfnamefont {N.~D.}\ \bibnamefont
  {Lanzillotti-Kimura}}, \bibinfo {author} {\bibfnamefont {A.}~\bibnamefont
  {Fainstein}}, \bibinfo {author} {\bibfnamefont {C.~A.}\ \bibnamefont
  {Balseiro}}, \ and\ \bibinfo {author} {\bibfnamefont {B.}~\bibnamefont
  {Jusserand}},\ }\href {\doibase 10.1103/PhysRevB.75.024301} {\bibfield
  {journal} {\bibinfo  {journal} {Phys. Rev. B}\ }\textbf {\bibinfo {volume}
  {75}},\ \bibinfo {pages} {024301} (\bibinfo {year} {2007})}\BibitemShut
  {NoStop}%
\bibitem [{\citenamefont {Lanzillotti-Kimura}\ \emph
  {et~al.}(2015)\citenamefont {Lanzillotti-Kimura}, \citenamefont {Fainstein},\
  and\ \citenamefont {Jusserand}}]{LanzillottiKimura201580}%
  \BibitemOpen
  \bibfield  {author} {\bibinfo {author} {\bibfnamefont {N.}~\bibnamefont
  {Lanzillotti-Kimura}}, \bibinfo {author} {\bibfnamefont {A.}~\bibnamefont
  {Fainstein}}, \ and\ \bibinfo {author} {\bibfnamefont {B.}~\bibnamefont
  {Jusserand}},\ }\href {\doibase https://doi.org/10.1016/j.ultras.2014.05.017}
  {\bibfield  {journal} {\bibinfo  {journal} {Ultrasonics}\ }\textbf {\bibinfo
  {volume} {56}},\ \bibinfo {pages} {80 } (\bibinfo {year} {2015})}\BibitemShut
  {NoStop}%
\bibitem [{\citenamefont {Rozas}\ \emph {et~al.}(2009)\citenamefont {Rozas},
  \citenamefont {Winter}, \citenamefont {Jusserand}, \citenamefont {Fainstein},
  \citenamefont {Perrin}, \citenamefont {Semenova},\ and\ \citenamefont
  {Lema\^{\i}tre}}]{PhysRevLett.102.015502}%
  \BibitemOpen
  \bibfield  {author} {\bibinfo {author} {\bibfnamefont {G.}~\bibnamefont
  {Rozas}}, \bibinfo {author} {\bibfnamefont {M.~F.~P.}\ \bibnamefont
  {Winter}}, \bibinfo {author} {\bibfnamefont {B.}~\bibnamefont {Jusserand}},
  \bibinfo {author} {\bibfnamefont {A.}~\bibnamefont {Fainstein}}, \bibinfo
  {author} {\bibfnamefont {B.}~\bibnamefont {Perrin}}, \bibinfo {author}
  {\bibfnamefont {E.}~\bibnamefont {Semenova}}, \ and\ \bibinfo {author}
  {\bibfnamefont {A.}~\bibnamefont {Lema\^{\i}tre}},\ }\href {\doibase
  10.1103/PhysRevLett.102.015502} {\bibfield  {journal} {\bibinfo  {journal}
  {Phys. Rev. Lett.}\ }\textbf {\bibinfo {volume} {102}},\ \bibinfo {pages}
  {015502} (\bibinfo {year} {2009})}\BibitemShut {NoStop}%
\bibitem [{\citenamefont {Soykal}\ \emph {et~al.}(2011)\citenamefont {Soykal},
  \citenamefont {Ruskov},\ and\ \citenamefont
  {Tahan}}]{PhysRevLett.107.235502}%
  \BibitemOpen
  \bibfield  {author} {\bibinfo {author} {\bibfnamefont {O.~O.}\ \bibnamefont
  {Soykal}}, \bibinfo {author} {\bibfnamefont {R.}~\bibnamefont {Ruskov}}, \
  and\ \bibinfo {author} {\bibfnamefont {C.}~\bibnamefont {Tahan}},\ }\href
  {\doibase 10.1103/PhysRevLett.107.235502} {\bibfield  {journal} {\bibinfo
  {journal} {Phys. Rev. Lett.}\ }\textbf {\bibinfo {volume} {107}},\ \bibinfo
  {pages} {235502} (\bibinfo {year} {2011})}\BibitemShut {NoStop}%
\bibitem [{\citenamefont {Fainstein}\ \emph {et~al.}(2013)\citenamefont
  {Fainstein}, \citenamefont {Lanzillotti-Kimura}, \citenamefont {Jusserand},\
  and\ \citenamefont {Perrin}}]{PhysRevLett.110.037403}%
  \BibitemOpen
  \bibfield  {author} {\bibinfo {author} {\bibfnamefont {A.}~\bibnamefont
  {Fainstein}}, \bibinfo {author} {\bibfnamefont {N.~D.}\ \bibnamefont
  {Lanzillotti-Kimura}}, \bibinfo {author} {\bibfnamefont {B.}~\bibnamefont
  {Jusserand}}, \ and\ \bibinfo {author} {\bibfnamefont {B.}~\bibnamefont
  {Perrin}},\ }\href {\doibase 10.1103/PhysRevLett.110.037403} {\bibfield
  {journal} {\bibinfo  {journal} {Phys. Rev. Lett.}\ }\textbf {\bibinfo
  {volume} {110}},\ \bibinfo {pages} {037403} (\bibinfo {year}
  {2013})}\BibitemShut {NoStop}%
\bibitem [{\citenamefont {Sauer}\ \emph {et~al.}(2010)\citenamefont {Sauer},
  \citenamefont {Daniels}, \citenamefont {Reiter}, \citenamefont {Kuhn},
  \citenamefont {Vagov},\ and\ \citenamefont {Axt}}]{PhysRevLett.105.157401}%
  \BibitemOpen
  \bibfield  {author} {\bibinfo {author} {\bibfnamefont {S.}~\bibnamefont
  {Sauer}}, \bibinfo {author} {\bibfnamefont {J.~M.}\ \bibnamefont {Daniels}},
  \bibinfo {author} {\bibfnamefont {D.~E.}\ \bibnamefont {Reiter}}, \bibinfo
  {author} {\bibfnamefont {T.}~\bibnamefont {Kuhn}}, \bibinfo {author}
  {\bibfnamefont {A.}~\bibnamefont {Vagov}}, \ and\ \bibinfo {author}
  {\bibfnamefont {V.~M.}\ \bibnamefont {Axt}},\ }\href {\doibase
  10.1103/PhysRevLett.105.157401} {\bibfield  {journal} {\bibinfo  {journal}
  {Phys. Rev. Lett.}\ }\textbf {\bibinfo {volume} {105}},\ \bibinfo {pages}
  {157401} (\bibinfo {year} {2010})}\BibitemShut {NoStop}%
\bibitem [{\citenamefont {Kr\"ugel}\ \emph {et~al.}(2006)\citenamefont
  {Kr\"ugel}, \citenamefont {Axt},\ and\ \citenamefont
  {Kuhn}}]{PhysRevB.73.035302}%
  \BibitemOpen
  \bibfield  {author} {\bibinfo {author} {\bibfnamefont {A.}~\bibnamefont
  {Kr\"ugel}}, \bibinfo {author} {\bibfnamefont {V.~M.}\ \bibnamefont {Axt}}, \
  and\ \bibinfo {author} {\bibfnamefont {T.}~\bibnamefont {Kuhn}},\ }\href
  {\doibase 10.1103/PhysRevB.73.035302} {\bibfield  {journal} {\bibinfo
  {journal} {Phys. Rev. B}\ }\textbf {\bibinfo {volume} {73}},\ \bibinfo
  {pages} {035302} (\bibinfo {year} {2006})}\BibitemShut {NoStop}%
\bibitem [{\citenamefont {{Wigger}}\ \emph {et~al.}(2017)\citenamefont
  {{Wigger}}, \citenamefont {{Czerniuk}}, \citenamefont {{Reiter}},
  \citenamefont {{Bayer}},\ and\ \citenamefont {{Kuhn}}}]{2017arXiv170104209W}%
  \BibitemOpen
  \bibfield  {author} {\bibinfo {author} {\bibfnamefont {D.}~\bibnamefont
  {{Wigger}}}, \bibinfo {author} {\bibfnamefont {T.}~\bibnamefont
  {{Czerniuk}}}, \bibinfo {author} {\bibfnamefont {D.~E.}\ \bibnamefont
  {{Reiter}}}, \bibinfo {author} {\bibfnamefont {M.}~\bibnamefont {{Bayer}}}, \
  and\ \bibinfo {author} {\bibfnamefont {T.}~\bibnamefont {{Kuhn}}},\
  }\href@noop {} {\bibfield  {journal} {\bibinfo  {journal} {ArXiv e-prints}\ }
  (\bibinfo {year} {2017})},\ \Eprint {http://arxiv.org/abs/1701.04209}
  {arXiv:1701.04209 [cond-mat.mes-hall]} \BibitemShut {NoStop}%
\bibitem [{\citenamefont {Czerniuk}\ \emph {et~al.}(2017)\citenamefont
  {Czerniuk}, \citenamefont {Wigger}, \citenamefont {Akimov}, \citenamefont
  {Schneider}, \citenamefont {Kamp}, \citenamefont {H\"ofling}, \citenamefont
  {Yakovlev}, \citenamefont {Kuhn}, \citenamefont {Reiter},\ and\ \citenamefont
  {Bayer}}]{PhysRevLett.118.133901}%
  \BibitemOpen
  \bibfield  {author} {\bibinfo {author} {\bibfnamefont {T.}~\bibnamefont
  {Czerniuk}}, \bibinfo {author} {\bibfnamefont {D.}~\bibnamefont {Wigger}},
  \bibinfo {author} {\bibfnamefont {A.~V.}\ \bibnamefont {Akimov}}, \bibinfo
  {author} {\bibfnamefont {C.}~\bibnamefont {Schneider}}, \bibinfo {author}
  {\bibfnamefont {M.}~\bibnamefont {Kamp}}, \bibinfo {author} {\bibfnamefont
  {S.}~\bibnamefont {H\"ofling}}, \bibinfo {author} {\bibfnamefont {D.~R.}\
  \bibnamefont {Yakovlev}}, \bibinfo {author} {\bibfnamefont {T.}~\bibnamefont
  {Kuhn}}, \bibinfo {author} {\bibfnamefont {D.~E.}\ \bibnamefont {Reiter}}, \
  and\ \bibinfo {author} {\bibfnamefont {M.}~\bibnamefont {Bayer}},\ }\href
  {\doibase 10.1103/PhysRevLett.118.133901} {\bibfield  {journal} {\bibinfo
  {journal} {Phys. Rev. Lett.}\ }\textbf {\bibinfo {volume} {118}},\ \bibinfo
  {pages} {133901} (\bibinfo {year} {2017})}\BibitemShut {NoStop}%
\bibitem [{\citenamefont {Tavis}\ and\ \citenamefont
  {Cummings}(1968)}]{PhysRev.170.379}%
  \BibitemOpen
  \bibfield  {author} {\bibinfo {author} {\bibfnamefont {M.}~\bibnamefont
  {Tavis}}\ and\ \bibinfo {author} {\bibfnamefont {F.~W.}\ \bibnamefont
  {Cummings}},\ }\href {\doibase 10.1103/PhysRev.170.379} {\bibfield  {journal}
  {\bibinfo  {journal} {Phys. Rev.}\ }\textbf {\bibinfo {volume} {170}},\
  \bibinfo {pages} {379} (\bibinfo {year} {1968})}\BibitemShut {NoStop}%
\bibitem [{\citenamefont {Richter}\ \emph {et~al.}(2015)\citenamefont
  {Richter}, \citenamefont {Gegg}, \citenamefont {Theuerholz},\ and\
  \citenamefont {Knorr}}]{PhysRevB.91.035306}%
  \BibitemOpen
  \bibfield  {author} {\bibinfo {author} {\bibfnamefont {M.}~\bibnamefont
  {Richter}}, \bibinfo {author} {\bibfnamefont {M.}~\bibnamefont {Gegg}},
  \bibinfo {author} {\bibfnamefont {T.~S.}\ \bibnamefont {Theuerholz}}, \ and\
  \bibinfo {author} {\bibfnamefont {A.}~\bibnamefont {Knorr}},\ }\href
  {\doibase 10.1103/PhysRevB.91.035306} {\bibfield  {journal} {\bibinfo
  {journal} {Phys. Rev. B}\ }\textbf {\bibinfo {volume} {91}},\ \bibinfo
  {pages} {035306} (\bibinfo {year} {2015})}\BibitemShut {NoStop}%
\bibitem [{\citenamefont {Dicke}(1954)}]{PhysRev.93.99}%
  \BibitemOpen
  \bibfield  {author} {\bibinfo {author} {\bibfnamefont {R.~H.}\ \bibnamefont
  {Dicke}},\ }\href {\doibase 10.1103/PhysRev.93.99} {\bibfield  {journal}
  {\bibinfo  {journal} {Phys. Rev.}\ }\textbf {\bibinfo {volume} {93}},\
  \bibinfo {pages} {99} (\bibinfo {year} {1954})}\BibitemShut {NoStop}%
\bibitem [{\citenamefont {Leymann}\ \emph {et~al.}(2015)\citenamefont
  {Leymann}, \citenamefont {Foerster}, \citenamefont {Jahnke}, \citenamefont
  {Wiersig},\ and\ \citenamefont {Gies}}]{PhysRevApplied.4.044018}%
  \BibitemOpen
  \bibfield  {author} {\bibinfo {author} {\bibfnamefont {H.~A.~M.}\
  \bibnamefont {Leymann}}, \bibinfo {author} {\bibfnamefont {A.}~\bibnamefont
  {Foerster}}, \bibinfo {author} {\bibfnamefont {F.}~\bibnamefont {Jahnke}},
  \bibinfo {author} {\bibfnamefont {J.}~\bibnamefont {Wiersig}}, \ and\
  \bibinfo {author} {\bibfnamefont {C.}~\bibnamefont {Gies}},\ }\href {\doibase
  10.1103/PhysRevApplied.4.044018} {\bibfield  {journal} {\bibinfo  {journal}
  {Phys. Rev. Applied}\ }\textbf {\bibinfo {volume} {4}},\ \bibinfo {pages}
  {044018} (\bibinfo {year} {2015})}\BibitemShut {NoStop}%
\bibitem [{\citenamefont {Kopylov}\ \emph {et~al.}(2015)\citenamefont
  {Kopylov}, \citenamefont {Radonji\ifmmode~\acute{c}\else \'{c}\fi{}},
  \citenamefont {Brandes}, \citenamefont {Bala\ifmmode~\check{z}\else
  \v{z}\fi{}},\ and\ \citenamefont {Pelster}}]{PhysRevA.92.063832}%
  \BibitemOpen
  \bibfield  {author} {\bibinfo {author} {\bibfnamefont {W.}~\bibnamefont
  {Kopylov}}, \bibinfo {author} {\bibfnamefont {M.}~\bibnamefont
  {Radonji\ifmmode~\acute{c}\else \'{c}\fi{}}}, \bibinfo {author}
  {\bibfnamefont {T.}~\bibnamefont {Brandes}}, \bibinfo {author} {\bibfnamefont
  {A.}~\bibnamefont {Bala\ifmmode~\check{z}\else \v{z}\fi{}}}, \ and\ \bibinfo
  {author} {\bibfnamefont {A.}~\bibnamefont {Pelster}},\ }\href {\doibase
  10.1103/PhysRevA.92.063832} {\bibfield  {journal} {\bibinfo  {journal} {Phys.
  Rev. A}\ }\textbf {\bibinfo {volume} {92}},\ \bibinfo {pages} {063832}
  (\bibinfo {year} {2015})}\BibitemShut {NoStop}%
\bibitem [{\citenamefont {Su}\ \emph {et~al.}(2013)\citenamefont {Su},
  \citenamefont {Bimberg}, \citenamefont {Knorr},\ and\ \citenamefont
  {Carmele}}]{PhysRevLett.110.113604}%
  \BibitemOpen
  \bibfield  {author} {\bibinfo {author} {\bibfnamefont {Y.}~\bibnamefont
  {Su}}, \bibinfo {author} {\bibfnamefont {D.}~\bibnamefont {Bimberg}},
  \bibinfo {author} {\bibfnamefont {A.}~\bibnamefont {Knorr}}, \ and\ \bibinfo
  {author} {\bibfnamefont {A.}~\bibnamefont {Carmele}},\ }\href {\doibase
  10.1103/PhysRevLett.110.113604} {\bibfield  {journal} {\bibinfo  {journal}
  {Phys. Rev. Lett.}\ }\textbf {\bibinfo {volume} {110}},\ \bibinfo {pages}
  {113604} (\bibinfo {year} {2013})}\BibitemShut {NoStop}%
\bibitem [{\citenamefont {Genway}\ \emph {et~al.}(2014)\citenamefont {Genway},
  \citenamefont {Li}, \citenamefont {Ates}, \citenamefont {Lanyon},\ and\
  \citenamefont {Lesanovsky}}]{PhysRevLett.112.023603}%
  \BibitemOpen
  \bibfield  {author} {\bibinfo {author} {\bibfnamefont {S.}~\bibnamefont
  {Genway}}, \bibinfo {author} {\bibfnamefont {W.}~\bibnamefont {Li}}, \bibinfo
  {author} {\bibfnamefont {C.}~\bibnamefont {Ates}}, \bibinfo {author}
  {\bibfnamefont {B.~P.}\ \bibnamefont {Lanyon}}, \ and\ \bibinfo {author}
  {\bibfnamefont {I.}~\bibnamefont {Lesanovsky}},\ }\href {\doibase
  10.1103/PhysRevLett.112.023603} {\bibfield  {journal} {\bibinfo  {journal}
  {Phys. Rev. Lett.}\ }\textbf {\bibinfo {volume} {112}},\ \bibinfo {pages}
  {023603} (\bibinfo {year} {2014})}\BibitemShut {NoStop}%
\bibitem [{\citenamefont {Ceban}\ \emph {et~al.}(2017)\citenamefont {Ceban},
  \citenamefont {Longo},\ and\ \citenamefont {Macovei}}]{PhysRevA.95.023806}%
  \BibitemOpen
  \bibfield  {author} {\bibinfo {author} {\bibfnamefont {V.}~\bibnamefont
  {Ceban}}, \bibinfo {author} {\bibfnamefont {P.}~\bibnamefont {Longo}}, \ and\
  \bibinfo {author} {\bibfnamefont {M.~A.}\ \bibnamefont {Macovei}},\ }\href
  {\doibase 10.1103/PhysRevA.95.023806} {\bibfield  {journal} {\bibinfo
  {journal} {Phys. Rev. A}\ }\textbf {\bibinfo {volume} {95}},\ \bibinfo
  {pages} {023806} (\bibinfo {year} {2017})}\BibitemShut {NoStop}%
\bibitem [{\citenamefont {Walter}\ \emph {et~al.}(2015)\citenamefont {Walter},
  \citenamefont {Nunnenkamp},\ and\ \citenamefont
  {Bruder}}]{ANDP:ANDP201400144}%
  \BibitemOpen
  \bibfield  {author} {\bibinfo {author} {\bibfnamefont {S.}~\bibnamefont
  {Walter}}, \bibinfo {author} {\bibfnamefont {A.}~\bibnamefont {Nunnenkamp}},
  \ and\ \bibinfo {author} {\bibfnamefont {C.}~\bibnamefont {Bruder}},\ }\href
  {\doibase 10.1002/andp.201400144} {\bibfield  {journal} {\bibinfo  {journal}
  {Annalen der Physik}\ }\textbf {\bibinfo {volume} {527}},\ \bibinfo {pages}
  {131} (\bibinfo {year} {2015})}\BibitemShut {NoStop}%
\bibitem [{\citenamefont {{Gegg}}\ \emph {et~al.}(2017)\citenamefont {{Gegg}},
  \citenamefont {{Carmele}}, \citenamefont {{Knorr}},\ and\ \citenamefont
  {{Richter}}}]{2017arXiv170502889G}%
  \BibitemOpen
  \bibfield  {author} {\bibinfo {author} {\bibfnamefont {M.}~\bibnamefont
  {{Gegg}}}, \bibinfo {author} {\bibfnamefont {A.}~\bibnamefont {{Carmele}}},
  \bibinfo {author} {\bibfnamefont {A.}~\bibnamefont {{Knorr}}}, \ and\
  \bibinfo {author} {\bibfnamefont {M.}~\bibnamefont {{Richter}}},\ }\href@noop
  {} {\bibfield  {journal} {\bibinfo  {journal} {ArXiv e-prints}\ } (\bibinfo
  {year} {2017})},\ \Eprint {http://arxiv.org/abs/1705.02889} {arXiv:1705.02889
  [quant-ph]} \BibitemShut {NoStop}%
\bibitem [{\citenamefont {{Hong}}\ \emph {et~al.}(2017)\citenamefont {{Hong}},
  \citenamefont {{Riedinger}}, \citenamefont {{Marinkovic}}, \citenamefont
  {{Wallucks}}, \citenamefont {{Hofer}}, \citenamefont {{Norte}}, \citenamefont
  {{Aspelmeyer}},\ and\ \citenamefont
  {{Gr{\"o}blacher}}}]{2017arXiv170603777H}%
  \BibitemOpen
  \bibfield  {author} {\bibinfo {author} {\bibfnamefont {S.}~\bibnamefont
  {{Hong}}}, \bibinfo {author} {\bibfnamefont {R.}~\bibnamefont {{Riedinger}}},
  \bibinfo {author} {\bibfnamefont {I.}~\bibnamefont {{Marinkovic}}}, \bibinfo
  {author} {\bibfnamefont {A.}~\bibnamefont {{Wallucks}}}, \bibinfo {author}
  {\bibfnamefont {S.~G.}\ \bibnamefont {{Hofer}}}, \bibinfo {author}
  {\bibfnamefont {R.~A.}\ \bibnamefont {{Norte}}}, \bibinfo {author}
  {\bibfnamefont {M.}~\bibnamefont {{Aspelmeyer}}}, \ and\ \bibinfo {author}
  {\bibfnamefont {S.}~\bibnamefont {{Gr{\"o}blacher}}},\ }\href@noop {}
  {\bibfield  {journal} {\bibinfo  {journal} {ArXiv e-prints}\ } (\bibinfo
  {year} {2017})},\ \Eprint {http://arxiv.org/abs/1706.03777} {arXiv:1706.03777
  [quant-ph]} \BibitemShut {NoStop}%
\bibitem [{\citenamefont {{Ding}}\ \emph {et~al.}(2017)\citenamefont {{Ding}},
  \citenamefont {{Yin}},\ and\ \citenamefont {{Li}}}]{2017arXiv170400446D}%
  \BibitemOpen
  \bibfield  {author} {\bibinfo {author} {\bibfnamefont {D.}~\bibnamefont
  {{Ding}}}, \bibinfo {author} {\bibfnamefont {X.}~\bibnamefont {{Yin}}}, \
  and\ \bibinfo {author} {\bibfnamefont {B.}~\bibnamefont {{Li}}},\ }\href@noop
  {} {\bibfield  {journal} {\bibinfo  {journal} {ArXiv e-prints}\ } (\bibinfo
  {year} {2017})},\ \Eprint {http://arxiv.org/abs/1704.00446} {arXiv:1704.00446
  [cond-mat.mes-hall]} \BibitemShut {NoStop}%
\bibitem [{\citenamefont {Bimberg}\ \emph {et~al.}(1999)\citenamefont
  {Bimberg}, \citenamefont {Grundmann},\ and\ \citenamefont
  {Ledentsov}}]{Bimberg:QD}%
  \BibitemOpen
  \bibfield  {author} {\bibinfo {author} {\bibfnamefont {D.}~\bibnamefont
  {Bimberg}}, \bibinfo {author} {\bibfnamefont {M.}~\bibnamefont {Grundmann}},
  \ and\ \bibinfo {author} {\bibfnamefont {N.~N.}\ \bibnamefont {Ledentsov}},\
  }\href@noop {} {\emph {\bibinfo {title} {Quantum Dot Heterostructures}}}\
  (\bibinfo  {publisher} {John Wiley \& Sons},\ \bibinfo {year}
  {1999})\BibitemShut {NoStop}%
\bibitem [{\citenamefont {Michler}(2003)}]{michler:QD}%
  \BibitemOpen
  \bibfield  {author} {\bibinfo {author} {\bibfnamefont {P.}~\bibnamefont
  {Michler}},\ }\href@noop {} {\emph {\bibinfo {title} {Single quantum dots:
  Fundamentals, applications and new concepts}}}\ (\bibinfo  {publisher}
  {Springer Science \& Business Media},\ \bibinfo {year} {2003})\BibitemShut
  {NoStop}%
\bibitem [{\citenamefont {Jahnke}(2012)}]{Jahnke:QD}%
  \BibitemOpen
  \bibfield  {author} {\bibinfo {author} {\bibfnamefont {F.}~\bibnamefont
  {Jahnke}},\ }\href@noop {} {\emph {\bibinfo {title} {Quantum optics with
  semiconductor nanostructures}}}\ (\bibinfo  {publisher} {Elsevier},\ \bibinfo
  {year} {2012})\BibitemShut {NoStop}%
\bibitem [{\citenamefont {Richter}\ \emph {et~al.}(2009)\citenamefont
  {Richter}, \citenamefont {Carmele}, \citenamefont {Butscher}, \citenamefont
  {Bücking}, \citenamefont {Milde}, \citenamefont {Kratzer}, \citenamefont
  {Scheffler},\ and\ \citenamefont {Knorr}}]{doi:10.1063/1.3117236}%
  \BibitemOpen
  \bibfield  {author} {\bibinfo {author} {\bibfnamefont {M.}~\bibnamefont
  {Richter}}, \bibinfo {author} {\bibfnamefont {A.}~\bibnamefont {Carmele}},
  \bibinfo {author} {\bibfnamefont {S.}~\bibnamefont {Butscher}}, \bibinfo
  {author} {\bibfnamefont {N.}~\bibnamefont {Bücking}}, \bibinfo {author}
  {\bibfnamefont {F.}~\bibnamefont {Milde}}, \bibinfo {author} {\bibfnamefont
  {P.}~\bibnamefont {Kratzer}}, \bibinfo {author} {\bibfnamefont
  {M.}~\bibnamefont {Scheffler}}, \ and\ \bibinfo {author} {\bibfnamefont
  {A.}~\bibnamefont {Knorr}},\ }\href {\doibase 10.1063/1.3117236} {\bibfield
  {journal} {\bibinfo  {journal} {Journal of Applied Physics}\ }\textbf
  {\bibinfo {volume} {105}},\ \bibinfo {pages} {122409} (\bibinfo {year}
  {2009})}\BibitemShut {NoStop}%
\bibitem [{\citenamefont {{Pleinert}}\ \emph {et~al.}(2017)\citenamefont
  {{Pleinert}}, \citenamefont {{von Zanthier}},\ and\ \citenamefont
  {{Agarwal}}}]{2017arXiv170205392P}%
  \BibitemOpen
  \bibfield  {author} {\bibinfo {author} {\bibfnamefont {M.-O.}\ \bibnamefont
  {{Pleinert}}}, \bibinfo {author} {\bibfnamefont {J.}~\bibnamefont {{von
  Zanthier}}}, \ and\ \bibinfo {author} {\bibfnamefont {G.~S.}\ \bibnamefont
  {{Agarwal}}},\ }\href@noop {} {\bibfield  {journal} {\bibinfo  {journal}
  {ArXiv e-prints}\ } (\bibinfo {year} {2017})},\ \Eprint
  {http://arxiv.org/abs/1702.05392} {arXiv:1702.05392 [quant-ph]} \BibitemShut
  {NoStop}%
\end{thebibliography}%
\end{document}